\documentclass[aps,pra,twocolumn,floatfix,superscriptaddress]{revtex4}  
\usepackage{float}
\usepackage{graphicx}
\usepackage{amssymb}
\usepackage{dsfont}
\usepackage{amsmath}  
\usepackage{epstopdf}
\usepackage{xcolor}
\begin{document}

\def\ket#1{|#1\rangle} 
\def\bra#1{\langle#1|}
\def\av#1{\langle#1\rangle}
\def\dkp#1{\kappa+i(\Delta+#1)}
\def\dkm#1{\kappa-i(\Delta+#1)}
\def\pp{{\prime\prime}}
\def\ppp{{\prime\prime\prime}}
\def\w{\omega}
\def\k{\kappa}
\def\D{\Delta}
\def\wp{\omega^\prime}
\def\wpp{\omega^{\prime\prime}}

\title{Recycling of a quantum field and optimal states for single-qubit rotations}
\author{Shanon Vuglar}
\affiliation{John Brown University, Siloam Springs, AR}
\author{Julio Gea-Banacloche}
\affiliation{University of Arkansas, Fayetteville, AR}

\date{\today}

\begin{abstract}
We introduce a family of quantized field states that can perform exact (entanglement- and error-free) rotations of a two-level atom starting from a specific state on the Bloch sphere.  We discuss the similarities and differences between these states and the recently-introduced ``transcoherent states.'' Our field states have the property that they are left unchanged after the rotation, and we find they are the asymptotic states obtained when a field interacts with a succession of identically prepared ancillary atoms.  Such a scheme was recently proposed [npj Quantum Information 3:17 (2017)] as a way to ``restore'' a field state after its interaction with a two-level atom, so as to reuse it afterwards, thus reducing the energy requirements for successive quantum logical operations. We generalize this scheme to find optimal pulses for arbitrary rotations, and also study analytically what happens if the ancillas are in a mixed, rather than a pure state.  Consistent with the numerical results in the original proposal, we find that as long as the ancilla preparation error is small (of the order of $1/\bar n$, where $\bar n$ is the average number of atoms in the pulses considered) it will introduce only higher-order errors in the performance of the restored pulse.    
\end{abstract}
\maketitle

\section{Introduction}

In general, the quantum nature of the radiation field will prevent the manipulation of the states of an atom with arbitrary accuracy, the ultimate constraints being set either by atom-field entanglement or by field fluctuations (or both).  The possible consequences of this for quantum information processing were first pointed out, independently, in \cite{jgb1,ozawa1}, and explored subsequently in a series of papers \cite{jgb2,gbozawa1,gbozawa2,karasawa}.  As a trivial example, if one tried to use a field in a number state $\ket n$ to perform what is commonly known as a $\pi/2$ pulse in quantum optics (a $\pi/2$ rotation on the Bloch sphere), starting from the ground state $\ket g$,
\begin{equation}
\ket g \to \frac{1}{\sqrt 2} \bigl( \ket g + e^{i\phi} \ket e \bigr)
\label{n1}
\end{equation}
(where $\ket e$ is the excited state, and $\phi$ an arbitrary phase), conservation of energy would instead produce an atom-field entangled state of the form
\begin{equation}
\ket g\ket n \to \frac{1}{\sqrt 2} \bigl( \ket g\ket n + e^{i\phi} \ket e \ket{n-1} \bigr).
\label{n2}
\end{equation}
The orthogonality of the field states in this expression means that the atomic state in (\ref{n2}) is totally mixed, and could be described as an incoherent superposition of the desired state, $\ket{\psi_\text{ideal}}$ (right-hand side of (\ref{n1})), and an orthogonal one.  Here, the error in the operation, which we can define as
\begin{equation}
\epsilon = 1- \bra{\psi_\text{ideal}} \rho^{(at)} \ket{\psi_\text{ideal}} = \frac 1 2 
\label{n3}
\end{equation}
(where $\rho^{(at)}$ is the reduced density matrix of the atom) is maximal, and the underlying cause is entanglement due to a conservation law, a point of view close to that of \cite{ozawa1}.  

If instead one had chosen to use a field in a coherent state $\ket{\alpha}$, where the initial number of photons is not sharply defined, one could in principle do much better, but the number and phase uncertainties of the coherent state would still have resulted in an error that scales as $1/\bar n$, with $\bar n = |\alpha|^2$ the average number of photons in the coherent state \cite{jgb1,koashi}:
\begin{equation}
\epsilon = \frac{\pi^2}{64}\left(\frac{\Delta n}{\bar n}\right)^2 + \frac{(\Delta \phi)^2}{4} = \frac{\pi^2}{64\bar n} +\frac{1}{16 \bar n},
\label{n4}
\end{equation} 
since in a coherent state $\Delta n = \sqrt{\bar n}$ and $\Delta\phi = 1/2\sqrt{\bar n}$.  A natural question one can ask at this point is whether there are other field states that can do even better than the coherent state.  This has been addressed recently in several papers from different groups \cite{ikonen,transc1,transc2}.  An important motivation is the question of the energy resources needed for quantum computation \cite{jgb2}, which has received renewed attention lately \cite{auffeves}.

In this context, the result established in \cite{gbozawa2}, that ``minimum energy pulses for quantum logic cannot be shared" is specially relevant.  The precise result established in \cite{gbozawa2} was that if a pulse containing  $\bar n$ photons was used to perform the same operation (specifically a $\pi/2$ pulse) on a set of $N$ atoms, at least for some initial atomic states a lower bound on the total error would be
\begin{equation}
\epsilon \ge \frac 1 4 \frac{N^2}{2N(N+1) + 4\bar n},
\end{equation}
and hence the error per atom, $\epsilon/N$, actually grows with the number of atoms $N$, at approximately a linear rate (although, in reality, sublinear) as long as $\bar n \gg N$ \cite{note}.  The implication is that the basic energy requirement of $\bar n$ photons to achieve a target error, $\epsilon$, per atom and operation, cannot be circumvented by using the same pulse multiple times: $N$ operations will still require a total energy that grows as $\sim N/\epsilon$.

Surprisingly, in \cite{ikonen} J. Ikonen, J. Salmilehto and M. M\" ott\" onen have proposed a scheme that seems to get rid of this requirement, at least in principle.  Their proposal involves ``cleaning up'' the control field after each interaction with a target atom so as to restore it to (something close to) its initial state; this would be accomplished by having it interact successively with a number of ancillary two-level atoms, prepared in a special superposition state. Perhaps the most remarkable aspect of their proposal is that preparing the ancillas, and restoring them to their appropriate initial state, could also be done with a single traveling pulse, which could, according to their numerical calculations, be reused for many rounds of ancilla preparation with negligible impact on the overall error rate.

Here we present a study of how the ``clean-up'' process works, first assuming the preparation of the ancillas is ``perfect'' at the beginning of every round.  We find that in this case the control field converges, after interacting with a sufficient number of ancillas, to a certain pure state (for which we give an analytical form), which actually minimizes the ``gate error,'' i.e., the average of the error over all possible initial states of the target atom, for a $\pi$-pulse operation.  We also show how by preparing the ancillas in different states one can obtain field states that minimize the gate error for other rotations.  Interestingly, we find that these are very similar to the ``transcoherent states'' introduced in \cite{transc1,transc2} and designed to produce Bloch-sphere rotations without entanglement when starting from a specific atomic state (typically the ground or excited state).  We discuss in detail the similarities and differences between our asymptotic states and the transcoherent states in Section III below.

Finally, and as a first attempt to understand the results obtained in \cite{ikonen} when a single itinerant pulse is used repeatedly to prepare and reset the ancillary atoms, we consider what happens when the initial state of the ancillas is not pure but a statistical mixture.  In that case, the asymptotic state of the field is mixed as well, but for small enough preparation errors (and large enough $\bar n$), we find that the state is approximately given by a mixture of two states, both of which perform the desired rotation with approximately the same error.  This means that, contrary to our expectations, but in general agreement with the numerical results of \cite{ikonen}, ``cleaning up'' the control pulse by having it interact with an ensemble of imperfectly-prepared ancillas does not increase the error of the operation performed by the control pulse substantially (relative to the perfect-ancilla case), at least as long as the ancilla preparation error remains sufficiently small itself.

\section{``cleaning up" the field: repeated interactions with identical ancillas }

\subsection{The $\pi$ rotation case}

As in \cite{ikonen,transc1,transc2}, all our results will be based on the Jaynes-Cummings model (JCM) \cite{jcm}, which describes the interaction of a two-level atom with a single-mode quantized field, in the rotating-wave approximation.  We'll write the Hamiltonian in the form
\begin{equation}
H = \hbar g \Bigl(a \ket e \bra g + a^\dagger \ket g\bra e \Bigr).
\label{e1}
\end{equation}
Pure states of the atom can be represented as points on the Bloch sphere \cite{optcom}, and a classical field could, by an appropriate choice of its phase and interaction time, perform arbitrary rotations on that sphere. For a quantum field, however, we find that after the interaction the field and atom are generally entangled, so neither of them is in a pure state.  In particular, the state of the field is modified by the interaction, so if it is used again an error in the intended rotation will result.  

In reference \cite{ikonen}, the authors proposed a way to ``clean up'' the field state by having it interact sequentially with a set of ancillary atoms, all prepared in the initial state 
\begin{equation}
\ket{\psi}_\text{ancilla} = \frac{1}{\sqrt{2}} \bigl(\ket g + i\ket e\bigr),
\label{e2}
\end{equation}
which corresponds to the point on the positive $y$ axis on the Bloch sphere. 
This assumes the initial field has zero phase (with the convention implicit in our choice of Hamiltonian (\ref{e1})). The initial average number of photons $n_0$ and the interaction time $T$ with each ancilla should be chosen so that
\begin{equation}
g\sqrt{n_0}\,T \simeq \pi/2.
\label{e3}
\end{equation}
This is the condition to cause a rotation of the state of the atom by $\pi$ radians around the $x$ axis (a ``$\pi$ pulse'').  As we shall see below, interaction with the ancillas asymptotically produces a pulse whose average photon number only satisfies Eq.~(\ref{e3}) approximately, yet it is optimal (in a sense to be made precise below) to perform a $\pi$ rotation in the interaction time given.  We shall also show, in Section II.B, how to prepare optimal pulses for other rotations, by preparing the ancillas in different initial states.

Given an initial product state of the system in the form $\ket{\Psi(0)} = (C_g\ket g + C_e\ket e)\otimes\ket{\Phi(0)}$, the state at time $t$ can be written, in matrix notation, in the $\{\ket g, \ket e\}$ basis, as follows 
\begin{equation}
\ket{\Psi(t)} = \begin{pmatrix} U_{gg} & U_{ge} \cr U_{eg} & U_{ee}\end{pmatrix} \begin{pmatrix} C_g(0) \cr C_e(0) \end{pmatrix}  \ket{\Phi(0)}
\label{e4}
\end{equation}
where the operators $U_{ij}(t)$ act on the field state and are given by
\begin{align}
U_{gg} &= \cos(gt\sqrt{a^\dagger a}) \cr
U_{ge} &= -i\sin(gt\sqrt{a^\dagger a})\frac{1}{\sqrt{a^\dagger a}}a^\dagger \cr
U_{eg} &= -i\sin(gt\sqrt{a a^\dagger})\frac{1}{\sqrt{a a^\dagger}}a \cr
U_{ee} &= \cos(gt\sqrt{a a^\dagger})
\label{e5}
\end{align}
(compare Eq.~(5) of \cite{optcom}, only noting that in that paper the atomic basis vectors were chosen in reverse order, $\{\ket e, \ket g\}$).  

Suppose the density matrix of the field after interaction with $n$ ancillas is $\rho_n^{(f)}$.  For the next interaction, the starting total density matrix can be written as
\begin{equation}
\rho_n = \rho_n^{(f)}\otimes \bigl(\ket\psi\bra\psi\bigr)_\text{ancilla} = \rho_n^{(f)}\otimes \begin{pmatrix} 1 & -i \cr i & 1 \end{pmatrix}
\end{equation}
where $\ket\psi_\text{ancilla}$ is given by (\ref{e2}).  After the interaction, the total density matrix $\rho_{n+1}$ will be
\begin{equation}
\rho_{n+1} =\frac 1 2 \begin{pmatrix} U_{gg} & U_{ge} \cr U_{eg} & U_{ee}\end{pmatrix}\rho_n^{(f)}\begin{pmatrix} 1 & -i \cr i & 1 \end{pmatrix} \begin{pmatrix} U_{gg}^\dagger & U_{eg}^\dagger \cr U_{ge}^\dagger & U_{ee}^\dagger\end{pmatrix}
\label{e6}
\end{equation}
where the $U_{ij}$ operators are evaluated at the time $T$ given by Eq.~(\ref{e3}).  Carrying out the multiplication and tracing over the ancilla, we obtain the new field state,
\begin{align}
\rho^{(f)}_{n+1} = &\frac 1 2\left(U_{ee}-iU_{eg}\right)\rho_n^{(f)}\left(U_{ee}^\dagger +i U_{eg}^\dagger\right) \cr
&+\frac 1 2 \left(U_{ge}-iU_{gg}\right)\rho_n^{(f)}\left(U_{ge}^\dagger +i U_{gg}^\dagger\right).
\label{e7}
\end{align}
A fixed point of this transformation would be provided by a pure field state that was a simultaneous eigenstate of 
$U_{ee}(T)-iU_{eg}(T)$ and $U_{ge}(T)-iU_{gg}(T)$ with eigenvalues of unit magnitude.  Such a state can formally be constructed.  Using the explicit expressions (\ref{e5}), it is easy to see that a state of the form $\ket{\Phi_\pi} = \sum_{n=0}^\infty C_n \ket n$, with coefficients $C_n$ satisfying the recursion relation 
\begin{equation}
C_{n+1} = \cot\left(\tfrac 1 2 gT\sqrt{n+1}\right) C_n
\label{e8}
\end{equation}
satisfies 
\begin{align}
\left(U_{ee}(T)-iU_{eg}(T)\right)\ket{\Phi_\pi} &= -\ket{\Phi_\pi}\cr
\left(U_{ge}(T)-iU_{gg}(T)\right)\ket{\Phi_\pi} &= -i\ket{\Phi_\pi}.
\label{e9}
\end{align}
The subscript $\pi$ anticipates that this state will have something to do with a $\pi$ pulse, and indeed it can be seen from Eqs.~(\ref{e9}) and (\ref{e4}) that, after interaction with the ancilla for a time $T$, the joint field-atom state is transformed as
\begin{equation}
\frac{1}{\sqrt{2}} \bigl(\ket g + i\ket e\bigr)\ket{\Phi_\pi} \to \frac{1}{\sqrt{2}} \bigl(\ket g - i\ket e\bigr)\ket{\Phi_\pi}.
\label{e10}
\end{equation}
That is, the state $\ket{\Phi_\pi}$ has the property that it causes an exact $\pi$ rotation of the ancilla state (\ref{e2}), leaving the global state unentangled and the field itself unchanged in the process.

It is, of course, not immediately obvious that the iteration (\ref{e7}), starting from an arbitrary field state, will necessarily converge to the pure state $\ket{\Phi_\pi}$, and there are also potential problems with the recursion relation (\ref{e8}): in particular, it would diverge if $\frac 1 2 gT\sqrt{n+1}$ was ever equal to an integer multiple of $\pi$ for some $n$.  

Because the state $\ket{\Phi_\pi}$ is a $\pi$ pulse, we expect its average number of photons, $\bar n$, to be close to $(\pi/2gT)^2$.  Anticipating things again, let us define
\begin{equation}
n_\Theta = \left(\frac{\Theta}{2gT}\right)^2
\label{e11}
\end{equation}
which will allow us to write the interaction time for a $\Theta$-rotation pulse in terms of a number $n_\Theta$ of the order of the average number of photons in the pulse.  In our case, for the $\pi$ pulse defined by the recursion relation (\ref{e8}), we'll write $gT = \pi/2\sqrt{n_\pi}$.  We can assume, without real loss of generality, that we choose $T$ so that $n_\pi$ is an integer, in which case the recursion relation terminates at $n_{max}= 4n_\pi -1$.  As this happens before the cotangent function has a chance to diverge (which would happen for $n = 16n_\pi -1$), we see that normalizable states $\ket{\Phi_\pi}$ can be found for any integer $n_\pi$.

\begin{figure}
    \includegraphics[width=9.2cm]{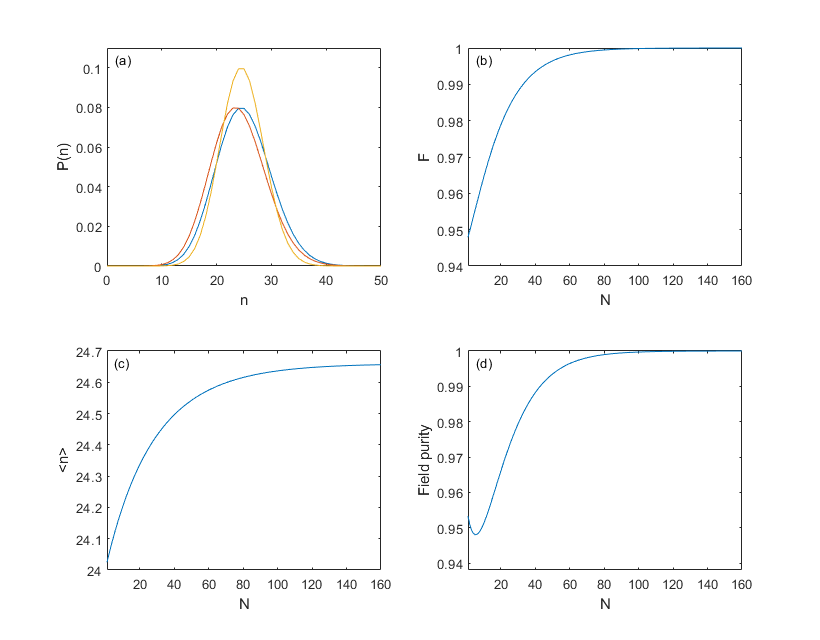}%
    {\caption{ (a) Photon number distribution for the initial field (blue, rightmost curve), the field after interaction with an atom initially in the ground state (red, leftmost curve), and the final field state (yellow, tallest, yellow curve) after interaction with a large number of ancillas. (b) Overlap between the final field state and the field state resulting from the interaction with $N$ ancillas.  (c) Average photon number in the field and (d) purity of the field as a function of the number of ancilla interactions, $N$.  }}
\label{fig:fig1}
\end{figure} 

To check the iteration (\ref{e7}) for convergence, we have carried out a series of numerical experiments.  Figure 1 shows what typically happens when the starting field is already quite close to the final state; it is also in the spirit of the ``field cleanup'' proposed in \cite{ikonen}.  For this calculation, we started with a coherent state $\ket\alpha$ with $\alpha =5$, i.e., $\bar n = 25$, and had it interact with an atom initially in the ground state for a time satisfying Eq.~(\ref{e3}), so the atom was transferred to the excited state and the field lost, on average, one photon.  We then take the resulting field state (which is slightly mixed and has $\av{n} = 24$) as the initial field for the interaction with the ancillas.  These are all prepared in the state (\ref{e2}); the interaction time with each ancilla is chosen to be  $gT = \pi/10$, i.e., we chose $n_\pi = 25$. The first graph shows the photon number distribution for the initial field state, the state after interacting with the atom in the ground state, and the final field state, which has $\av{n} \simeq 24.7$ and near unit purity after interacting with about 50 ancillas.  It also has a slightly narrower number distribution than the initial state, reflecting the fact that it is slightly amplitude-squeezed.  These features can be derived analytically, as we will show below.

\begin{figure}
    \includegraphics[width=9.2cm]{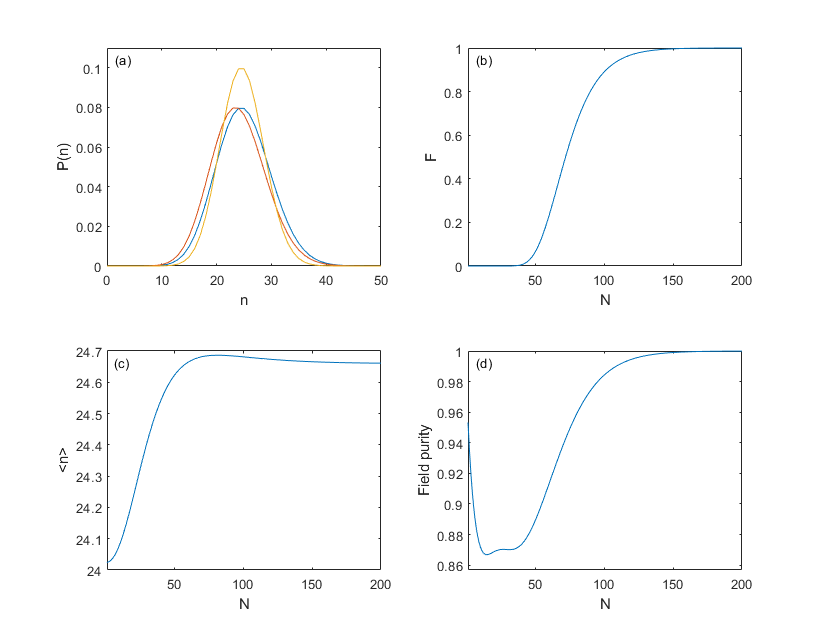}%
    {\caption{As in Fig.~1, but the initial field state has the opposite phase to the one in Fig.~1.}}
\label{fig:fig2}
\end{figure} 

To verify that the final field state is essentially independent of the initial one, we show, in Figure 2, the same calculation but starting from a $\ket{\Phi(0)}$ that is nearly orthogonal to $\ket{\Phi_\pi}$.  Specifically, instead of starting the field in a coherent state $\ket\alpha$ with $\alpha = 5$ we now choose $\alpha = -5$, i.e., a field $180^\circ$ out of phase with the one we used as the initial state in Fig.~1.  In spite of this, the field eventually evolves to the same final state as in Fig. 1, as shown by the fidelity (overlap between $\ket{\Phi_\pi}$ and the field after $N$ ancilla iterations) plotted in Fig. 2(b). Only, the convergence is somewhat slower (the fidelity starts from practically zero in this case), and the initial drop in purity is somewhat more pronounced than in Fig. 1(d).  

A useful analytical approximation to the state $\ket{\Phi_\pi}$ can be obtained as follows. First, writing $C_{n+1} \simeq C(n) + dC/dn$, we can turn Eq.~(\ref{e8}) into a differential equation:
\begin{equation}
\frac{dC}{dn} = \left[\cot\left(\frac{\pi}{4\sqrt{n_\pi}}\,\sqrt{n+1}\right) - 1\right] C(n).
\label{e12}
\end{equation}
This can be integrated formally to yield
\begin{equation}
C(n) =C(0) \exp\left(\int_0^n \left[\cot\left(\frac{\pi}{4\sqrt{n_\pi}}\,\sqrt{n^\prime+1}\right) - 1\right] dn^\prime \right).
\label{e13}
\end{equation}
To get a Gaussian approximation, one can expand the integrand around $n^\prime = n_\pi-1$, giving 
\begin{equation}
 \cot\left(\frac{\pi}{4\sqrt{n_\pi}}\,\sqrt{n^\prime+1}\right) - 1 \simeq -\frac{\pi}{4 n_\pi} (n+1-n_\pi) 
 \label{e14}
\end{equation}
and therefore (after normalization)
\begin{equation}
C(n)\simeq \frac{1}{\sqrt{2\sqrt{n_\pi}}} \,e^{-(\pi/8 n_\pi)(n+1-n_\pi)^2}.
\label{e15}
\end{equation}
The state (\ref{e15}) has average number of photons $\bar n = n_\pi -1$, and a width $\Delta n = \sqrt{2 n_\pi/\pi}$.  The latter is a good approximation to the results of our numerical experiments, but the former is slightly off.  

We can improve on the approximation (\ref{e15}) by noting that, when $n_\pi$ is an integer, $\cot[(\pi/4 n_\pi)\sqrt{n+1}] =1$ for $n=n_\pi -1$, so (by the recursion formula (\ref{e8})), $C_{n_\pi} = C_{n_\pi-1}$.  This means the peak of the function $C(n)$, and hence the point where $dC/dn=0$, should be approximately at $n=n_\pi -\frac 1 2$.  We can then replace the approximation (\ref{e12}) by   
\begin{equation}
\frac{dC}{dn} = \left[\cot\left(\frac{\pi}{4\sqrt{n_\pi}}\,\sqrt{n+\tfrac 1 2}\right) - 1\right] C(n)
\label{e16}
\end{equation}
and proceed with the integration as before, this time expanding the integrand around $n=n_\pi -\frac 1 2$.  We also keep one more term in the expansion, to account for the slight asymmetry observed in the numerical experiments.  The result is
\begin{align}
C(n)\simeq &\frac{1}{\sqrt{2\sqrt{n_\pi}}} \,e^{-(\pi/8 n_\pi)(n+1/2-n_\pi)^2} \cr
&\quad \times\left(1+\frac{\pi(2+\pi)}{96 n_\pi^2} \left(n+1/2-n_\pi\right)^3 \right).
\label{e17}
\end{align}
Now the average number of photons is approximately given by
\begin{equation}
\bar n \simeq n_\pi -\frac 1 2 +\frac{2+\pi}{8\sqrt{n_\pi}}
\label{e18}
\end{equation}
which agrees quite well with numerical results (e.g., for $n_\pi = 25$ we obtain $\bar n = 24.66$, while Eq.~(\ref{e18}) gives $\bar n = 24.63$).  The width of the distribution, on the other hand, is still given to a good approximation by
\begin{equation}
\Delta n \simeq \sqrt{\frac{2 n_\pi}{\pi}} \simeq \sqrt{\frac{2 \bar n}{\pi}}
\label{e19}
\end{equation}
since the difference between the two expressions above is of order $1/\sqrt{\bar n}$.  

We note that (\ref{e19}) implies that the state $\ket{\Phi_\pi}$ has the optimal number squeezing required, according to \cite{ikonen}, to minimize the error in a $\pi$ pulse rotation, when the gate fidelity is averaged over all the initial atomic states.  

Figure 3 explores this further by comparing the action of the state $\ket{\Phi_\pi}$ to that of the ``standard'' squeezed states (i.e., states obtained by a displacement of the squeezed vacuum), for ensembles of 256 random initial atomic states, uniformly distributed on the Bloch sphere (a different ensemble was used for each value of the squeezing parameter $r$).  The black diamond shows the minimum of the mean error curve, which, for this particular numerical experiment,  is all but indistinguishable from the mean error achieved with the state $\ket{\Phi_\pi}$ (red diamond). 

\begin{figure}
    \includegraphics[width=9.2cm]{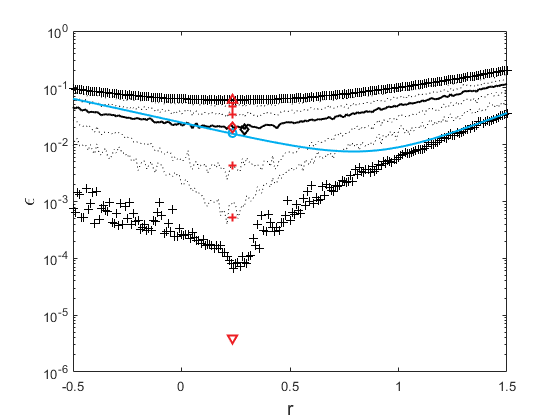}%
    {\caption{In black: minimum and maximum errors ($+$ symbols) obtained over ensembles of 256 atoms in random initial states when a squeezed field with squeeze parameter $r$ was used to perform a $\pi$ rotation; the mean (average) error for each set (continuous line), and the minimum of this curve (diamond); and the 10, 25, 75 and 90 percentiles (dotted lines).  In red: extreme error values obtained for a field in the state $\ket{\Phi_\pi}$ (triangles), the mean for this set (diamond), and  the corresponding percentiles ($+$ symbols).  In cyan: the error when the atom is initially in the ground state, for a squeezed field (solid line), and for the state $\ket{\Phi_\pi}$ (solid circle). }}
\label{fig:fig4}
\end{figure} 

Figure 3 also shows explicitly the gate error when the initial atomic state is the ground state $\ket g$, as a blue (cyan) line for the squeezed states, and as a blue circle for the state $\ket{\Phi_\pi}$.  Interestingly, this error is very close to the mean for the state $\ket{\Phi_\pi}$, but can be still reduced substantially by increasing the squeezing, albeit at the cost of increasing the average error.  This is relevant to the discussion of transcoherent states in Section III below.

\subsection{Arbitrary rotation case}

The nature of the field state produced by the  iteration process described in the previous section was, in our numerical experiments, determined exclusively by the choice (\ref{e2}) of the atomic ancilla's state.  As the fixed point of the iteration process, this final field state must be left invariant by the interaction with the ancilla.  Since the ancilla state is on the equator of the Bloch sphere, with equal amplitudes for the excited and ground states, it makes sense that the rotation caused by the field should correspond to a $\pi$ pulse, since that would take the atom to another equatorial state, without causing a change in the number of photons in the field.  

From this we can conjecture that if we carry out a similar iteration starting from a different ancilla state, such as 
\begin{equation}
\ket{\psi_i} = \cos\frac\theta 2 \ket g + e^{i\phi}\sin\frac\theta 2 \ket e
\label{e20}
\end{equation}
(where $\theta$ and $\phi$ are the colatitude and azimuth of the point representing the state in the Bloch sphere), the final field state will also produce a rotation of the Bloch vector such that the excited and ground state probabilities do not change.  This is accomplished by a rotation by an angle $\Theta = 2\theta$ in a plane containing both the $z$ axis and the initial Bloch vector, starting on one side of the $z$ axis and ending (with the same colatitude) on the other side (see Figure 4).
In what follows we show that this is indeed the case.

\begin{figure}
    \includegraphics[width=7cm]{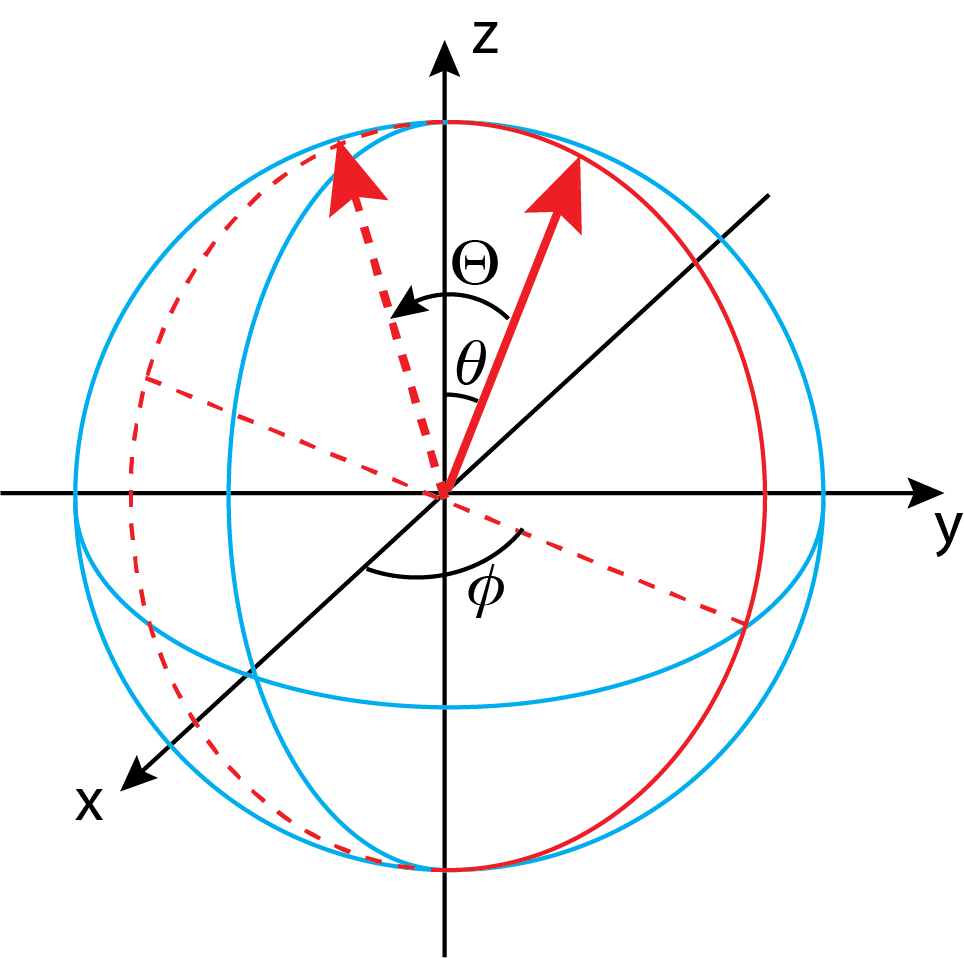}%
    {\caption{The rotation in the Bloch sphere produced by the state $\ket{\Phi_\Theta}$, when the initial ancilla state is given by Eq.~(\ref{e20}).  The rotation takes place in the vertical plane whose intersection with the $x$-$y$ plane is shown by a dotted straight line; the axis of rotation (not shown) would lie on the $x$-$y$ plane, at an angle of $\pi/2-\phi$ to the left of the $x$ axis.}}
\label{fig:fig4}
\end{figure} 

If, after interacting for a time $T$ each with $n$ atoms in the initial state (\ref{e20}), the field's density operator is $\rho_n^{(f)}$, after interaction with the $n+1$ atom the atom-field density matrix will be (compare Eq.~(\ref{e6}))
\begin{equation}
\rho_{n+1} = \begin{pmatrix} U_{gg} & U_{ge} \cr U_{eg} & U_{ee}\end{pmatrix}\rho_n^{(f)}\begin{pmatrix} c^2 & e^{-i\phi} s c \cr e^{i\phi} s c & s^2 \end{pmatrix} \begin{pmatrix} U_{gg}^\dagger & U_{eg}^\dagger \cr U_{ge}^\dagger & U_{ee}^\dagger\end{pmatrix}
\label{e21}
\end{equation}
where $c\equiv \cos(\theta/2)$ and $s\equiv\sin(\theta/2)$ have been introduced for convenience.  The recursion relation for the field density matrix, generalizing Eq.~(\ref{e7}), is
\begin{align}
\rho^{(f)}_{n+1} = &\left(s e^{i\phi}U_{ee}+c U_{eg}\right)\rho_n^{(f)}\left(s e^{-i\phi}U_{ee}^\dagger+c U_{eg}^\dagger\right) \cr
&+ \left(s e^{i\phi}U_{ge}+cU_{gg}\right)\rho_n^{(f)}\left(s e^{-i\phi}U_{ge}^\dagger +c U_{gg}^\dagger\right).
\label{e22}
\end{align}
The fixed point of this transformation is now the pure state $\rho^{(f)} = \ket{\Phi_\Theta}\bra{\Phi_\Theta}$ (with $\Theta = 2 \theta$), with $\ket{\Phi_\Theta}$ simultaneously satisfying
\begin{align}
\left(\cos\frac\theta 2 U_{eg} + e^{i\phi}\sin\frac\theta 2 U_{ee}\right)\ket{\Phi_\Theta} &= -e^{i\phi}\sin\frac\theta 2 \ket{\Phi_\Theta}\cr
\left(\cos\frac\theta 2 U_{gg} + e^{i\phi}\sin\frac\theta 2 U_{ge}\right)\ket{\Phi_\Theta} &= \cos\frac\theta 2 \ket{\Phi_\Theta}.
\label{e23}
\end{align}
It is easy to verify that the coefficients of $\ket{\Phi_\Theta}$ are given by the recursion relation
\begin{equation}
C_{n+1} = -i e^{i\phi} \tan\left(\frac\theta 2\right) \cot\left(\frac 1 2 gT\sqrt{n+1}\right) C_n.
\label{e24}
\end{equation}
Comparing this to Eq.~(\ref{e8}), we see that it also terminates at $n_{max}=(\pi/gT)^2-1$ (independently of $\theta$), but that the peak will be around the value $n_\Theta$ defined by Eq.~(\ref{e11}), with $\Theta = 2\theta$.  For small $\Theta$, therefore, $n_{max}$ can be much greater that $n_\Theta$; however, we have found numerically that the coefficients $C_n$ become completely negligible long before one reaches $n=n_{max}$ (something that also follows from the approximate results (\ref{e26}) and (\ref{e27}) below).

Using (\ref{e4}), (\ref{e20}) and (\ref{e23}), the effect of letting the state $\ket{\Phi_\Theta}$ interact with the state (\ref{e20}) for the time $T$ is to produce the transformation

\begin{widetext}
\begin{equation}
\left( \cos\frac\theta 2 \ket g + e^{i\phi}\sin\frac\theta 2 \ket e \right)\ket{\Phi_\Theta} \to \left( \cos\frac\theta 2 \ket g - e^{i\phi}\sin\frac\theta 2 \ket e \right)\ket{\Phi_\Theta}
\label{e25}
\end{equation}
which corresponds to the rotation described above, and shown in Figure 4.

An approximate expression for the state $\ket{\Phi_\Theta}$ can be obtained along the same lines as in the previous section.  Each coefficient $C_n$ consists of a factor $e^{in(\phi-\pi/2)}$, times
\begin{equation}
|C_n| \simeq \left(\frac{\Theta}{4\pi n_\Theta \sin(\Theta/2)}\right )^{1/4} e^{-(n-n_\Theta + 1/2)\Theta/8n_\Theta\sin(\Theta/2)}\left(1+\frac{\Theta(2+\Theta\cot(\Theta/2)}{96 n_\Theta^2 \sin(\Theta/2)}\left(n-n_\Theta+1/2 \right)^3\right)
\label{e26}
\end{equation}
\end{widetext}
with $n_\Theta$ defined as in Eq.~(\ref{e11}).  The width of the distribution is approximately
\begin{equation}
\Delta n \simeq \sqrt{\frac{2 \bar n \sin(\Theta/2)}{\Theta}}.
\label{e27}
\end{equation}
According to \cite{transc2}, this is the optimal number squeezing to minimize the error for a $\Theta$ rotation, when averaged over all the initial states in the Bloch sphere, so the states $\ket{\Phi_\Theta}$ are optimal in that sense.  In particular, for the case of a $\pi/2$ rotation, we find $\Delta n = \sqrt{2 \sqrt 2 \bar n/\pi}$, which, according to \cite{ikonen}, is the optimum squeezing for a $\pi/2$-pulse. 

This gives us a possible way to prepare optimal field states to perform arbitrary rotations in the Bloch sphere. One needs to start from a set of ancillas prepared in a state like (\ref{e20}), with $\theta = \Theta/2$, and a field in, e.g., a coherent state. Successive interactions, each lasting for a time $T$, with the atoms in the set eventually will generate a field state that will rotate the ancilla state exactly, without entanglement and without changing itself, and will rotate all other initial atomic states with minimal error, on average.  (The effect of small errors in the preparation of the ancillas is considered in Section IV below.)

\section{Similarities and differences with transcoherent states}

In \cite{transc1}, Goldberg and Steinberg introduced a set of field states they named ``transcoherent,'' whose purpose was to perform an exact $\pi/2$ atomic rotation, starting from either the ground state or the excited state.  More recently, in \cite{transc2}, these states were generalized to perform arbitrary $\Theta$ rotations, always starting from either $\ket g$ or $\ket e$.  The transcoherent states are defined by recursion relations similar (but not identical) to (\ref{e24}), and also have Gaussian approximations similar to Eq.~(\ref{e26}) in the limit of large photon numbers.

There are two main differences between the transcoherent states and our $\ket{\Phi_\Theta}$ states.  First, the transcoherent states are designed to perform an exact rotation by an angle $\Theta$ starting from the ground or excited state, whereas for the states $\ket{\Phi_\Theta}$ the rotation is only exact when starting from a state of the form (\ref{e20}), with $\theta = \Theta/2$ (see also Fig.~4).  Second, when starting from the special state (\ref{e20}), our $\ket{\Phi_\Theta}$ are left unchanged by the interaction with the atom, whereas the transcoherent states in general are not (although, as shown in \cite{transc1}, they can still be reused after the interaction with an advantage over coherent states).

Another important difference, which follows from the above, is that the transcoherent states are only optimal for the special initial atomic state for which they are designed (as they rotate that state with zero error), whereas our $\ket{\Phi_\Theta}$ states are also optimal ``on average'' for arbitrary initial atomic states, as discussed at the end of the previous section.  In particular, the squeezing of the transcoherent states is given by \cite{transc2}
\begin{equation}
\Delta n_\text{transc} = \sqrt{\frac{\bar n \sin\Theta}{\Theta}}
\label{e28}
\end{equation}
which is not optimal in general, and in particular means that one cannot actually make a transcoherent state for a $\pi$ rotation, unlike our $\ket{\Phi_\pi}$ of Section II.A.

\begin{figure}
    \includegraphics[width=9.2cm]{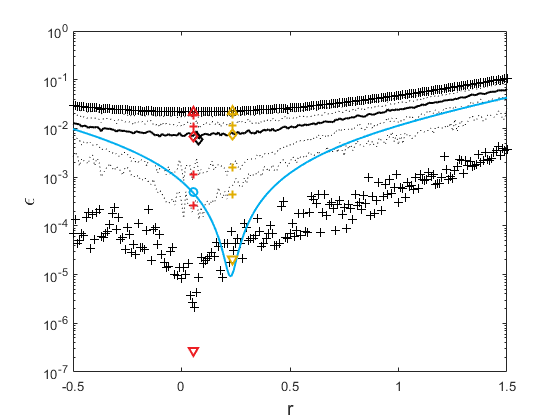}%
    {\caption{In black: minimum and maximum errors ($+$ symbols) obtained over an ensemble of 256 atoms in random initial states when a squeezed field with squeeze parameter $r$ was used to perform a $\pi/2$ rotation; the mean (average) error for each set (continuous line), and the minimum of this curve (diamond); and the 10, 25, 75 and 90 percentiles (dotted lines).  In red and dark yellow, respectively: extreme error values obtained for, respectively, the field state $\ket{\Phi_{\pi/2}}$ and a transcoherent state (triangles), the means for the corresponding set (diamond), and  the corresponding percentiles ($+$ symbols).  In cyan: error when the atom is initially in the ground state, for a squeezed field (solid line) and for the state $\ket{\Phi_{\pi/2}}$ (circle). }}
\label{fig:fig5}
\end{figure} 

Figure 5 illustrates some of these differences for the case of a $\pi/2$ rotation.  Both the state $\ket{\Phi_{\pi/2}}$ and the transcoherent state perform about as well as a squeezed state with the same degree of squeezing on a random atomic state, while clearly exceeding the squeezed state performance for special initial atomic states,  either very close to $\ket g$ (for the transcoherent state), or very close to the state (\ref{e20}) with $\theta = \pi/4$ (for the state $\ket{\Phi_{\pi/2}}$).  Of course, for exactly those initial atomic states, the error would go to zero and fall outside the range of the figure, but evidently our random sample of initial atomic states did not happen to include those states.  This explains why the minimum of the cyan solid curve, which is calculated for an initial atomic state $\ket g$ and a squeezed field state, happens to be lower than the minimum   error calculated for the transcoherent state over the random ensemble (dark yellow triangle).  

If we leave aside those exceptional initial atomic states, for which perfect (error-free) rotations are possible, and consider instead the average error over a random ensemble of initial atomic states, we note  that, for this particular numerical experiment, this is minimized by a squeezed field state (black diamond) whose degree of squeezing is pretty close to that of the state $\ket{\Phi_{\pi/2}}$ (red diamond).  More generally, the figure shows that when acting on a random sample of initial atomic states the average performance of both the transcoherent and the $\ket{\Phi_{\pi/2}}$ states is very close to that of an ordinary squeezed state with the same degree of squeezing.  This holds not only for the mean error but also, approximately, for the various percentiles shown in Fig.~5 (and, for the state $\ket{\Phi_{\pi}}$, in Fig.~3 as well).  In other words, while these special field states presumably can perform rotations with very small error starting from some initial atomic states (those close to either Eq.~(\ref{e20}), for the state $\ket{\Phi_{2\theta}}$, or to $\ket g$ or $\ket e$ for a transcoherent state), the improvement over an ordinary squeezed state seems to be appreciable only for a small subset of the atomic states sampled, perhaps only about $10\%$ or less.

It is probably fair to say that each of these kinds of field states will be useful in different situations: if the initial atomic state is known to be $\ket g$ or $\ket e$, then the transcoherent states will be optimal, whereas if the initial state is completely unknown our $\ket{\Phi_\Theta}$ (or, alternatively, a squeezed state with the same degree of squeezing) will minimize the average error.  Still other possibilities (initial state with known azimuth) are discussed in \cite{transc2}.


\section{Imperfectly-prepared ancillas}



In Section II we assumed the ancilla state (\ref{e2}) or (\ref{e20}) was prepared exactly, but, of course, if a pulse with a finite number of photons is used, this is unlikely to be the case.  Assuming the ancillas start in the ground state, one might, in view of \cite{transc1} and   \cite{transc2}, consider using the appropriate transcoherent state, but this field will be changed by the interaction, and if reused for the next ancilla, some error will be inevitable.  

In this section we wish to study analytically what happens to a field that interacts sequentially with an ensemble of ancillas that, because of preparation errors, is described by a \emph{mixed state} density operator $\rho$. Our main result is that, if the density operator has the diagonal form
\begin{equation}
\rho_a = (1-\lambda)\ket{v_1}\bra{v_1} + \lambda \ket{v_1}_\perp\bra{v_1}_\perp
\label{e29}
\end{equation}
for orthogonal states $\ket{v_1}$ and $\ket{v_1}_\perp$ and $\lambda \ll 1$, the fixed point of the iteration process 
\begin{equation}
\rho_{n+1}^{(f)} = {\rm Tr}_a\,\bigl [ U \rho_{n}^{(f)}\otimes\rho_a U^\dagger \bigr]
\label{e30}
\end{equation}
is a field described approximately (to lowest order in $\lambda$, and in $1/\bar n$, the average number of photons) by the density matrix
\begin{equation}
\rho^{(f)} = (1-\lambda)\ket{\Phi_1}\bra{\Phi_1} + \lambda \ket{\Phi_2}\bra{\Phi_2}.
\label{e31}
\end{equation}
Here, $\ket{\Phi_1}$ is the field state that would result from the repeated interaction with an ensemble of ancillas in the pure state $\ket{v_1}$, whereas $\ket{\Phi_2}$ is a state orthogonal to $\ket{\Phi_1}$.  Assuming, without loss of generality, that the state $\ket{v_1}$ has the form (\ref{e20}), an explicit expression for the state $\ket{\Phi_2}$ is
\begin{align}
\ket{\Phi_2} &= \frac{1}{\sqrt{{\cal N}_2}} \left(cs U_{gg}-e^{i\phi}c^2 U_{ge}-e^{-i\phi}s^2 U_{eg}+sc U_{ee}\right)\ket{\Phi_1}\cr
&=\frac{1}{\sqrt{{\cal N}_2} }\left[ \frac s c (1+U_{ee}) - \frac c s (1-U_{gg})\right] \ket{\Phi_1}
\label{e32}
\end{align}
where $c\equiv \cos(\theta/2)$ and $s\equiv\sin(\theta/2)$, and ${\cal N}_2$ is a normalization constant.  For the specific case of a $\pi$ pulse, with $\theta = \phi = \pi/2$, we have
\begin{align}
\ket{\Phi_2} &=\frac{1}{\sqrt{{\cal N}_2} }\left( U_{ee} + U_{gg}\right) \ket{\Phi_1} \cr
&\simeq \sqrt{\frac{\pi}{2 n_\pi}} \sum_{n=0}^\infty \left(n-n_\pi+\frac 1 2 \right) C_n \ket n
\label{e33}
\end{align}
where $C_n$ are the coefficients of the state $\ket{\Phi_\pi}$, given approximately by Eq.~(\ref{e17}), and $n_\pi$, assumed large, is given by (\ref{e11}) with $\Theta = \pi$.  (The proofs of the results (\ref{e31}) and (\ref{e32}) are given in the Appendix.)

If the field state $\ket{\Phi_1}$, with an average number of photons $\bar n \simeq n_\pi$, is used to try to perform a $\pi$ rotation on an atom, the error, averaged over all possible initial states, will be $\epsilon_1 \simeq \pi/6 \bar n$.  If instead the field state $\ket{\Phi_2}$ is used, the error (which can be calculated from the above expressions and the explicit forms (\ref{e5}), with $gt = \pi/2\sqrt{\bar n}$) will be $\epsilon_2 = 7\pi/16 \bar n$.  Hence, if the mixed state of the field (\ref{e31}) is used, the error will be $\epsilon_1$ with probability $1-\lambda$, and $\epsilon_2$ with probability $\lambda$, for a total error of
\begin{equation}
\epsilon =\epsilon_1 + \lambda(\epsilon_2-\epsilon_1) \simeq \frac{\pi}{6 \bar n} + \lambda \,\frac{13\pi}{48\bar n}.
\label{e34}
\end{equation}
Recall that $\lambda$ is the preparation error of the ancillas, so it too should be expected to scale as $1/\bar n$.  In \cite{ikonen}, the authors proposed a scheme where all the ancillas are prepared and reset by a single itinerant pulse.  Presumably this leads to an error that grows with the number of ancillas, but as long as this number is less than $\bar n$ Eq.~(\ref{e34}) suggests that the ``clean-up'' process for the field will be reasonably effective; that is, one should be able to reuse the field at least a few times to perform $\pi$ rotations on the target qubits with close to minimal error, as indeed the numerical simulations of \cite{ikonen} suggest.

Perhaps the most surprising takeaway from the result (\ref{e34}) is that \emph{lack of purity in the field state does not necessarily translate into a gate error}.  The field state (\ref{e31}) is mixed, an incoherent superposition of two orthogonal states, but, as it happens, both states can accomplish the desired task (the $\pi$ rotation of the qubit), because they have approximately the same number of photons, relatively well defined intensity and phase, and would interact with the atom for the same length of time.  Under those circumstances, the lack of purity (characterized by $\lambda$ in the equations above) only matters when it comes to weighing in the relative gate error for each of the two fields: as Eq.~(\ref{e34}) shows, if $\epsilon_1$ and $\epsilon_2$ were the same, $\lambda$ would not even contribute to the total gate error.  

Essentially, this means that prior entanglement of the field with something else is not necessarily a source of error for a gate operation.  As already suggested by our discussion in Section III, very different kinds of field states can yield essentially the same average error, as long as they have the same amplitude and phase fluctuations.  From this perspective, the main problem with reusing a pulse is that prior interaction will typically increase $\Delta n$, because of the possibility of gaining or losing a photon in the process.  In the recycling scheme of \cite{ikonen}, this is a more pronounced risk factor for the main pulse (because it is expected to interact with atoms that may be close to the ground or the excited state) than for the pulse being used to reset the ancillas (which only need to be taken from one near-equatorial state to another); this may be, ultimately, why   the latter pulse can be used to ``clean up'' the former.

\section{Conclusions}
We have presented here an analytical treatment of the field ``clean-up'' protocol proposed in \cite{ikonen}, and showed that, for the case of perfectly prepared ancillas, the ``clean-up'' process actually results in the preparation of a field state that has a number of similarities to the ``transcoherent states'' introduced in \cite{transc1,transc2}.  In particular, the field states we find have, like the transcoherent states, the property of being able to carry out an atomic state rotation with zero error and no field-atom entanglement, for some particular initial atomic state.  In addition, and unlike the transcoherent states, the field states introduced here minimize the average error of the operation over all the initial atomic states.

We have also considered to some extent the question of what happens when an initial field state interacts repeatedly with imperfectly-prepared ancillas, and found the somewhat surprising result that, even though this reduces the purity of the resulting field state, it does not necessarily increase the gate error substantially. More precisely, as Eq.~(\ref{e34}) shows, the increase in the gate error scales as the ancilla preparation error divided by the average number of photons in the pulse, so if the ancilla preparation error  ($\lambda$ in Eq.~(\ref{e34}))  also scales as $1/\bar n$, the gate error (which already scales as $1/\bar n$  in the perfect ancilla case) only increases by an amount of order $1/\bar n^2$.  This, we believe, goes some way towards explaining what, in our opinion, is the most remarkable result in \cite{ikonen}, namely, the numerical calculations showing that a single preparation pulse could be reused over and over to prepare the ancillas, without a substantial impact on the gate error (or rather, with an error that, initially at least, increases very slowly with the number of reuses). 

To conclude, we should comment on the main limitation of all these proposals (including our own), which is their reliance on the Jaynes-Cummings model, i.e., on a single-mode treatment of the field, an approximation that can only really hold for an atom strongly coupled to a resonant cavity. For an atom, or a set of atoms, in free space, the spontaneous emission coupling to the vacuum modes is naturally stronger than the coherent coupling to the driving laser mode, and this results in a larger error than the ones considered in either \cite{ikonen} or \cite{transc1,transc2}, albeit one that still scales as $1/\bar n$ (since that is the ratio of the spontaneous to the stimulated emission rate) \cite{reply}.  On the other hand, as already suggested in \cite{ikonen}, it might be possible to approach the error limits considered here for atoms (real or artificial) strongly confined in one-dimensional geometries, such as those found in waveguide or circuit quantum electrodynamics \cite{cqed}.  We note that for some of these systems, there are already theoretical proposals for logical gates requiring only extremely low powers, at the single-photon level (see, e.g., \cite{koshino,brod2}). However, these typically work in the adiabatic regime and require long interaction times, with the error scaling as some power of $1/T$ (e.g., $1/T^2$ in \cite{leno1}); moreover, they may also involve ``hidden'' energy costs for photon routing.  This last possibility makes the suggestion in \cite{ikonen}, that some of these techniques might find application for photon routers or circulators (such as the systems considered in \cite{routers}), particularly intriguing.  Still, we anticipate that a full temporal multimode treatment of the field will be necessary to ascertain the range of applicability of the present results to more realistic systems.  

\begin{acknowledgements}
We acknowledge the MonArk NSF Quantum Foundry supported by the National Science Foundation Q-AMASE-i program under NSF award No. DMR-1906383.
\end{acknowledgements}

\appendix

\section{Ancilla in a mixed state}

To prove the results (\ref{e31}) and (\ref{e32}) of the main text, we will first establish the following properties of the states $\ket{\Phi_\Theta}$:

Let $\ket{\Phi_1}\equiv\ket{\Phi_\Theta}$ be the field state that generates an exact rotation when starting from the atomic state
\begin{equation}
\ket{v_1} = \cos\frac\theta 2 \ket g + e^{i\phi}\sin\frac\theta 2 \ket e
\label{a1}
\end{equation}
with $\Theta = 2\theta$.  Let $\ket{v_1^\prime}$ be the rotated state (as in the right-hand side of (\ref{e25})); let also $\ket{v_2} \equiv \ket{v_1}_\perp$ be the state orthogonal to $\ket{v_1}$ (opposite point on the Bloch sphere), $\ket{v_2^\prime}$ the result of applying a $\Theta$ rotation to $\ket{v_2}$, and $\ket{v_2^\prime}_\perp \equiv \ket{v_1^\prime}$ the state orthogonal to $\ket{v_2^\prime}$. Explicitly, we have
\begin{align}
\ket{v_1^\prime} &= \cos\frac\theta 2 \ket g - e^{i\phi}\sin\frac\theta 2 \ket e \cr
\ket{v_2} &= \sin\frac\theta 2 \ket g -e^{i\phi} \cos\frac\theta 2 \ket e \cr
\ket{v_2^\prime} &= \sin\frac\theta 2 \ket g + e^{i\phi}\cos\frac\theta 2 \ket e 
\label{a2}
\end{align}
We then find the following: under the evolution producing the rotation $\Theta$, that is, such that 
\begin{equation}
U\ket{\Phi_1}\ket{v_1} = \ket{\Phi_1}\ket{v_1^\prime},
\label{a3}
\end{equation}
we have
\begin{align}
U\ket{\Phi_1}\ket{v_2} &\simeq -\left(1-\frac{\theta\sin\theta}{2n_\Theta}\right)\ket{\Phi_1}\ket{v_2^\prime} + \sqrt{\frac{\theta\sin\theta}{n_\Theta}}\ket{\Phi_2}\ket{{v_1^\prime}} \cr
U\ket{\Phi_2}\ket{v_1} &\simeq \left(1-\frac{\theta\sin\theta}{2n_\Theta}\right)\ket{\Phi_2}\ket{v_1^\prime} + \sqrt{\frac{\theta\sin\theta}{n_\Theta}}\ket{\Phi_1}\ket{{v_2^\prime}} 
\label{a4}
\end{align}
where $\ket{\Phi_2}$ is defined by Eq.~(\ref{e32}). It is easy to see that $\ket{\Phi_2}$ is orthogonal to $\ket{\Phi_1}$, if we observe that $U_{gg}$ and $U_{ee}$ are Hermitian, and $U_{eg}^\dagger = -U_{ge}$.  We then have 
\begin{align}
\av{\Phi_2|\Phi_1} &\propto \left(cs U_{gg}-e^{-i\phi}c^2 U_{ge}^\dagger-e^{i\phi}s^2 U_{eg}^\dagger+sc U_{ee}\right) \ket{\Phi_1}\cr
&=  \left(cs U_{gg}+e^{i\phi}c^2 U_{eg}+e^{-i\phi}s^2 U_{ge}+sc U_{ee}\right) \ket{\Phi_1}\cr
&=0
\label{a5}
\end{align}
by Eqs.~(\ref{e23}).

Besides the equations (\ref{a3}) and (\ref{a4}), an equation for the evolution of the initial state $\ket{\Phi_2}\ket{v_2}$ could also be written in detail, but for our purposes it is only necessary to observe that it, too, will lead to the rotation of $\ket{v_2}$, except for corrections of the order of $1/n_\Theta$:
\begin{equation}
U\ket{\Phi_2}\ket{v_2} \simeq \ket{\Phi_2}\ket{v_2^\prime} + \text{terms with norm of order } \frac{1}{n_\Theta}.
\label{a6}
\end{equation}
Eqs.~(\ref{a3}), (\ref{a4}),  and (\ref{a6}) show that, through order $1/n_\Theta$, only two field states are involved in the time evolution of $\rho^{(f)}\otimes\rho_a$, when these operators have the forms (\ref{e29}) and (\ref{e31}); moreover, in the outcome of each rotation each of the field states ends up associated with an orthogonal atomic state.  Let us then write $\rho^{(f)}$ in the form (\ref{e31}), only with $\lambda$ replaced by an $x$ to be determined later:
\begin{equation}
\rho^{(f)} = (1-x)\ket{\Phi_1}\bra{\Phi_1} + x \ket{\Phi_2}\bra{\Phi_2}.
\label{a7}
\end{equation}
\begin{widetext}
Assuming both $\lambda$ and $x$ to be small, and writing $\epsilon=\sqrt{\theta\sin\theta/n_\Theta}$ for brevity, we have the following evolution:
\begin{align}
U \rho^{(f)}\otimes\rho_a U^\dagger = \,&(1-\lambda)(1-x)\ket{\Phi_1}\bra{\Phi_1}\otimes\ket{v_1^\prime}\bra{v_1^\prime} \cr
&+\lambda(1-x)\Bigl[-(1-\tfrac 12\epsilon^2)\ket{\Phi_1}\ket{v_2^\prime} + \epsilon\ket{\Phi_2}\ket{{v_1^\prime}} \Bigr] \Bigl[-(1-\tfrac 12\epsilon^2)\bra{\Phi_1}\bra{v_2^\prime} + \epsilon\bra{\Phi_2}\bra{v_1^\prime} \Bigr] \cr
&+(1-\lambda)x\Bigl[(1-\tfrac 12\epsilon^2)\ket{\Phi_2}\ket{v_1^\prime} - \epsilon\ket{\Phi_1}\ket{v_2^\prime} \Bigr] \Bigl[(1-\tfrac 12\epsilon^2)\bra{\Phi_2}\bra{v_1^\prime} + \epsilon\bra{\Phi_1}\bra{v_2^\prime}\Bigr] \cr
&+\lambda x\Bigl[\ket{\Phi_2}\bra{\Phi_2}\otimes\ket{v_2^\prime}\bra{v_2^\prime} + O(1/n_\Theta)\Bigr].
\end{align}
\label{a8}
Taking the trace over the atomic system, and recalling $\ket{v_1^\prime}$ and $\ket{v_2^\prime}$ are orthogonal, we  get
\begin{equation}
{\rm Tr}_a\,\bigl [ U \rho^{(f)}\otimes\rho_a U^\dagger \bigr] = \left[1-x+\epsilon^2(x-\lambda)\right]\ket{\Phi_1}\bra{\Phi_1} + \left[x+\epsilon^2(\lambda-x)\right]\ket{\Phi_2}\bra{\Phi_2}.
\label{a9}
\end{equation}
Letting $x=\lambda$ we recover Eq.~(\ref{e31}), and so $\rho^{(f)}$ as given by Eq.~(\ref{e31}) is, indeed, the fixed point of the recursion relation (\ref{e30}).

It remains only to prove the results (\ref{a4}) and (\ref{a6}).  The result (\ref{a6}) is more or less straightforward, because if the time $gt$ is chosen appropriately and the field in the operators (\ref{e5}) is treated classically, any initial state will be rotated by the angle $\Theta = 2gt\sqrt{n_\Theta}$, and the difference between treating the field classically or quantum-mechanically, to lowest order, will always be terms with a norm (squared) of the order of $1/\bar n$. 

Turning then to the first of Eqs.~(\ref{a4}), we start with the explicit expressions (\ref{a2}) for the atomic states involved, and the usual representation of the evolution operator $U$ in terms of the field operators $U_{ij}$, to write
\begin{equation}
U\ket{\Phi_1}\ket{v_2} = \left(s^2 U_{gg} - sc e^{i\phi} U_{ge} + sc e^{-i\phi} U_{eg} -c^2 U_{ee}\right)\ket{\Phi_1}\ket{v_2^\prime} + \left(cs U_{gg} - e^{i\phi} c^2 U_{ge}- e^{-i\phi} s^2 U_{eg} + sc U_{ee}\right)\ket{\Phi_1}\ket{v_1^\prime}
\label{a10}
\end{equation}
with the usual shorthand $s=\sin(\theta/2)$, $c=\cos(\theta/2)$.  The field state in the second term is, by definition, $\ket{\Phi_2}$, up to a normalization constant that we will calculate presently.  The first term can be simplified if we use the identities in (\ref{e23}) to eliminate the parts containing the operators $U_{eg}$ and $U_{ge}$, which allows us to write 
\begin{equation}
\left(s^2 U_{gg} - sc e^{i\phi} U_{ge} + sc e^{-i\phi} U_{eg} -c^2 U_{ee}\right)\ket{\Phi_1} = -\ket{\Phi_1} +\left(U_{gg}-U_{ee}\right)\ket{\Phi_1}.
\label{a11}
\end{equation}
\end{widetext}

Now it is easy to calculate $\bra{\Phi_1}(U_{gg}-U_{ee})\ket{\Phi_1}$ to the relevant order, which is the lowest nonvanishing order.  Since for a $\Theta = 2\theta$ pulse we want $gt = \theta/\sqrt{n_\Theta}$, where $n_\Theta$ is approximately the center of the photon number distribution, expanding around that point we have
\begin{align}
U_{gg} &= \cos\theta-\frac{\theta\sin\theta}{2 n_\Theta}(\hat n -n_\Theta) \cr
U_{ee} &= \cos\theta-\frac{\theta\sin\theta}{2 n_\Theta}(\hat n +1-n_\Theta)
\label{a12}
\end{align}
where $\hat n$ is the photon number operator, from which it follows immediately that 
\begin{equation}
 \bra{\Phi_1} \left(U_{gg} - U_{ee}\right)\ket{\Phi_1} = \frac{\theta\sin\theta}{2 n_\Theta}.
\label{a13}
\end{equation}
Putting this together with (\ref{a11}), we can see that Eq.~(\ref{a10}) can be written as
\begin{equation}
U\ket{\Phi_1}\ket{v_2} \simeq -\left(1-\frac{\theta\sin\theta}{2n_\Theta}\right)\ket{\Phi_1}\ket{v_2^\prime} + \sqrt{{\cal N}_2}\ket{\Phi_2}\ket{{v_1^\prime}} 
\label{a14}
\end{equation}
plus possibly other terms, proportional to $\ket{v_2^\prime}$ and involving contributions from other field states; however, such terms are actually negligible to this order, since, as we shall see, the norm of the second term is just what is needed for the expression (\ref{a14}) to be already normalized to order $1/n_\Theta$. 

To see this, and calculate ${\cal N}_2$ explicitly, we once again use Eqs.~(\ref{e23}) to eliminate the $U_{eg}$ and $U_{ge}$ terms, which yields the second line of Eq.~(\ref{e32}).  Substitution of the expansion (\ref{a12}) followed by some trigonometric manipulation then yields
\begin{align}
{{\cal N}_2}\av{\Phi_2|\Phi_2} &=  \frac{\theta^2}{n_\Theta^2}\bra{\Phi_1} \left(\hat n-n + \sin^2\left(\frac\theta 2\right) \right)^2\ket{\Phi_1} \cr
&\simeq  \frac{\theta^2}{n_\Theta^2} \,\frac{n_\Theta \sin\theta}{\theta} = \frac{\theta \sin\theta}{n_\Theta}
\label{a15}
\end{align}
since the expectation value on the first line is, to this order, essentially the variance of the photon number distribution in the state $\ket{\Phi_1}$, which we can get from Eq.~(\ref{e27}) in the main text.  This shows, as claimed, that Eq.~(\ref{a14}) is normalized to this order, and that the state $\ket{\Phi_2}$ will be normalized if we choose  ${\cal N}_2 = \theta\sin\theta/n_\Theta$.

To prove the second of Eqs.~(\ref{a4}), we start with the equivalent of Eq.~(\ref{a10}) for an ancilla initially in state $\ket{v_1}$ and field in state $\ket{\Phi_2}$:
\begin{widetext}
\begin{equation}
U\ket{\Phi_2}\ket{v_1} = \left(c^2 U_{gg} + sc e^{i\phi} U_{ge} -sc e^{-i\phi} U_{eg} -s^2 U_{ee}\right)\ket{\Phi_2}\ket{v_1^\prime} +  \left(sc U_{gg} + s^2 e^{i\phi} U_{ge}+c^2 e^{-i\phi} U_{ge} + sc U_{ee}\right)\ket{\Phi_2}\ket{ v_2^\prime}.
\label{a16}
\end{equation}
Now we want the projection of the first term onto $\ket{\Phi_2}$, and of the second term onto $\ket{\Phi_1}$.  For the first one, we note we can write
\begin{align}
&\left(c^2 U_{gg} + sc e^{i\phi} U_{ge} -sc e^{-i\phi} U_{eg} -s^2 U_{ee}\right)\ket{\Phi_2} = \frac{1}{\sqrt{{\cal N}_2}}\left(c^2 U_{gg} + sc e^{i\phi} U_{ge} -sc e^{-i\phi} U_{eg} -s^2 U_{ee}\right)\left[ \frac s c (1+U_{ee}) - \frac c s (1-U_{gg})\right] \ket{\Phi_1} \cr
&= \frac{1}{\sqrt{{\cal N}_2}}\left[ \frac s c (1+U_{ee}) - \frac c s (1-U_{gg})\right]\left(c^2 U_{gg} + sc e^{i\phi} U_{ge} -sc e^{-i\phi} U_{eg} -s^2 U_{ee}\right) \ket{\Phi_1} + \frac{sc}{\sqrt{{\cal N}_2}}\left[ e^{i\phi} U_{ge} - e^{-i\phi} U_{eg}, \frac s c U_{ee}+ \frac c s U_{gg}\right]  \ket{\Phi_1}   \cr
&= \frac{1}{\sqrt{{\cal N}_2}}\left[ \frac s c (1+U_{ee}) - \frac c s (1-U_{gg})\right] \ket{\Phi_1} + \frac{sc}{\sqrt{{\cal N}_2}}\left[ e^{i\phi} U_{ge} - e^{-i\phi} U_{eg}, \frac s c U_{ee}+ \frac c s U_{gg}\right]  \ket{\Phi_1} \cr
&= \ket{\Phi_2} + \frac{sc}{\sqrt{{\cal N}_2}}\left[ e^{i\phi} U_{ge} - e^{-i\phi} U_{eg}, \frac s c U_{ee}+ \frac c s U_{gg}\right]  \ket{\Phi_1} 
\label{a17}
\end{align}
\end{widetext}
where we have used the results (\ref{e23}) to simplify the second line, so now we only have to evaluate the commutator shown.  By calculating matrix elements in the $\ket n$ basis, and expanding again the relevant operators, we obtain the result
\begin{align}
&\left[ e^{i\phi} U_{ge} - e^{-i\phi} U_{eg}, \frac s c U_{ee}+ \frac c s U_{gg}\right] \cr
&\quad\simeq -i\frac{\theta\sin\theta}{n_\pi} \left(e^{i\phi} \ket{n+1}\bra n + e^{-i\phi} \ket n\bra{n+1} \right).\cr
\label{a18}
\end{align}
We can then use this and the specific expressions for the coefficients of $\ket{\Phi_1}$ and $\ket{\Phi_2}$ to calculate the projection of the second term of (\ref{a17}) onto $\ket{\Phi_2}$, keeping in mind that, around the maximum of the photon number distribution, the coefficients of $\ket{\Phi_1}$ satisfy approximately $C_{n+1}\simeq -i e^{i\phi} C_n$.  The result is simply 
\begin{align}
&\frac{sc}{\sqrt{{\cal N}_2}}\bra{\Phi_2}\left[ e^{i\phi} U_{ge} - e^{-i\phi} U_{eg}, \frac s c U_{ee}+ \frac c s U_{gg}\right]\ket{\Phi_1} \cr 
&\quad\simeq -\frac{\theta\sin\theta}{2n_\pi}.
\label{a19}
\end{align}
The second state in (\ref{a16}) can be manipulated in a similar way, using again Eqs.~(\ref{e23}) to show that 
\begin{align}
&\left(sc U_{gg} + s^2 e^{i\phi} U_{ge}+c^2 e^{-i\phi} U_{ge} + sc U_{ee}\right)\ket{\Phi_2} = 0 \cr
&\quad+  \frac{1}{\sqrt{{\cal N}_2}}\left[ s^2 e^{i\phi} U_{ge} + c^2 e^{-i\phi} U_{eg}, \frac s c U_{ee}+ \frac c s U_{gg}\right] \ket{\Phi_1}.
\label{a20}
\end{align}
Expanding the commutator as before, and projecting onto $\ket{\Phi_1}$ this time, we get
\begin{align}
&\bra{\Phi_1}\left(sc U_{gg} + s^2 e^{i\phi} U_{ge}+c^2 e^{-i\phi} U_{ge} + sc U_{ee}\right)\ket{\Phi_2}  \cr
&\quad\simeq  \sqrt{\frac{\theta\sin\theta}{n_\pi}}.
\label{a21}
\end{align}
Taking (\ref{a16}), (\ref{a17}), {\ref{a19}) and (\ref{a21}) together, the proof of the second of Eqs.~(\ref{a4}) is complete.

\end{document}